\documentclass[12pt,preprint]{aastex}
\usepackage{emulateapj5}                                                      \begin{document}
\newcommand{\myemail}{ian.george@gsfc.nasa.gov}
\shorttitle{Circumnuclear X-rays from NGC~3516}
\shortauthors{George et al.}
\title{The Detection of Circumnuclear X-ray Emission 
from the Seyfert Galaxy NGC~3516}

\author {I.M. George\altaffilmark{1,2},
T.J. Turner\altaffilmark{1,2},
H. Netzer\altaffilmark{3},
S.B. Kraemer\altaffilmark{4,5},
J. Ruiz\altaffilmark{4,5},
D. Chelouche\altaffilmark{3},
D.M. Crenshaw\altaffilmark{6},
T. Yaqoob\altaffilmark{2,7},
K. Nandra\altaffilmark{2,8},
R.F. Mushotzky\altaffilmark{2}
}

\altaffiltext{1}{Joint Center for Astrophysics, 
        Department of Physics,
	University of Maryland, Baltimore County, 
	1000 Hilltop Circle, Baltimore, MD 21250}
\altaffiltext{2}{Laboratory for High Energy Astrophysics, Code 662,
        NASA/Goddard Space Flight Center,
        Greenbelt, MD 20771}
\altaffiltext{3}{School of Physics and Astronomy and the Wise Observatory,
        The Beverly and Raymond Sackler Faculty of Exact Sciences,
        Tel Aviv University, Tel Aviv 69978, Israel.}
\altaffiltext{4}{Laboratory for Astronomy and Solar Physics, Code 681,
         NASA/Goddard Space Flight Center,
        Greenbelt, MD 20771}
\altaffiltext{5}{Institute for Astrophysics and Computational Sciences, 
        The Catholic University of America, Washington D.C. 20064}
\altaffiltext{6} {Department of Physics and Astronomy, 
	Georgia State University, Atlanta, GA 30303}
\altaffiltext{7}{Department of Physics \& Astronomy, 
	Johns Hopkins University, 3400 North Charles Street, 
	Baltimore, MD 21218}
\altaffiltext{8}{Universities Space Research Association}

\slugcomment{Accepted for publication in {\em The Astrophysical Journal}}

\begin{abstract}
We present the first high-resolution, X-ray image of the circumnuclear 
regions of the Seyfert 1 galaxy NGC~3516, 
using the {\it Chandra X-ray Observatory} ({\it CXO}).
All three of the {\it CXO} observations reported were performed 
with one of the two grating assemblies in place, and
here we restrict our analysis to undispersed photons 
(i.e. those detected in the zeroth-order).
A previously-unknown X-ray source is detected
$\sim$6~arcsec ($1.1 h_{75}^{-1}$~kpc)
NNE of the nucleus (position angle $\sim$29~degrees) 
which we designate 
CXOU~110648.1+723412. Its spectrum can be characterized 
as a power law with a photon index $\Gamma \sim$1.8--2.6, or 
as thermal emission with a temperature 
$kT \sim$0.7--3~keV. Assuming a location within NGC~3516,
isotropic emission implies a luminosity  
$L \sim$2--8$\times10^{39}h_{75}^{-2}\ {\rm erg\ s^{-1}}$
in the 0.4--2~keV band.
If due to a single point source, the object is 
super-Eddington for a $1.4 M_{\odot}$ neutron star.
However, multiple sources or a small, extended source 
cannot be excluded using the current data.
Large-scale extended X-ray emission is also detected out to 
$\sim 10$~arcsec ($\sim 1.7 h_{75}^{-1}$~kpc) 
from the nucleus to the NE and SW, and is approximately aligned 
with the morphologies of the radio emission and 
extended narrow emission line region (ENLR).
The mean luminosity of this emission is 
1--5$\times 10^{37}h_{75}^{-2}\ {\rm erg\ s^{-1}\ arcsec^{-2}}$, 
in the 0.4--2~keV band. 
Unfortunately the current data cannot usefully constrain 
its spectrum.
These results are consistent with earlier suggestions of 
circumnuclear X-ray emission in NGC~3516. 
If the extended emission is due to scattering of the nuclear X-ray 
continuum, then the pressure in the X-ray emitting gas 
is at least two orders of magnitude too small to 
provide the confining medium for the ENLR clouds. 
\end{abstract}

\keywords{
galaxies: active -- 
galaxies: individual (NGC~3516) --
galaxies: nuclei -- 
galaxies: Seyfert --
X-rays: galaxies}

\section{INTRODUCTION}
\label{sec:intro}

\begin{table*}
\begin{center}
\caption{{\it CXO} OBSERVING LOG}
\label{tab:obslog}
\begin{tabular}{cllcccc}
\hline
\hline
\multicolumn{7}{l}{\footnotesize \mbox{}}\\
Instrument 	&
	Seq./Obs.id. 	&
	\multicolumn{1}{c}{Start Time} 	&
	Duration 	&
	Roll$^{a}$	&
	$\Delta$RA $\cos(dec)^{b}$	&
	$\Delta$dec$^{b}$
\\
	&
	&
	\multicolumn{1}{c}{(UTC)}&
	(ks) 	&
	(degrees)	&
	(arcsec)	&
	(arcsec)
\\
        (1)   &
        (2)   &
        (3)   &
        (4)   &
        (5)   &
        (6)   &
        (7)   
\\
\multicolumn{7}{l}{\mbox{}}\\
\hline
\multicolumn{7}{l}{\mbox{}}\\
LETGS	&
	700136/0831	&
	2000 Sep 30 21:05:22	&
	47.0    &
	\phn21.3    &
	$+0.51$	&
	$+0.34$
\\
HETGS	&
	700270/2431	&
	2001 Apr 09 14:12:05 &
	39.9    &
	213.0    &
	$-0.41$	&
	$+0.26$
\\
HETGS	&
	700270/2080    &
	2001 Apr 10 17:55:54 & 
	75.4    &
	213.0    &
	$-0.43$	&
	$+0.23$
\\
\multicolumn{7}{l}{\mbox{}}
\\
\hline
\multicolumn{7}{l}{NOTES -- (${a}$) spacecraft roll angle, 
measured from North towards West;
(${b}$) X-ray position minus optical }\\
\multicolumn{7}{l}{position 
(RA=11h~06m~47.490s, dec=+72d~34m~06.88s; Clements 1981).}
\end{tabular}

\end{center}
\end{table*}

\noindent
NGC~3516 is an apparently undisturbed SB0 type, 
with the bar $\sim$25~arcsec long aligned N--S, plus  
a possible outer oval 30--100~arcsec with a 
major axis aligned in the NE--SW direction
(e.g. Adams 1977).
It is one of the ``nebulae'' included in the studies of 
Hubble (1926), and noted as having high-excitation emission lines
by Seyfert (1943).
The nucleus has since been shown to possess a 
Seyfert 1 spectrum (e.g. Khachikian \& Weedman 1974), and
since been studied extensively at all accessible wavebands from radio to X-rays
(e.g. see Arribas et al 1997; Guainazzi, Marshall, Parmar 2001;
and references therein).

Following Ferruit et al. (1998, and references therein), 
hereafter we adopt a distance of $35.7 h_{75}^{-1}$~Mpc 
to the nucleus of NGC~3516 
(where $h_{75} = H_0/75\ {\rm km\ s^{-1}\ Mpc^{-1}}$), 
thus 1~arcsec $\equiv 173 h_{75}^{-1}$~pc. 
In the radio, an unresolved, flat-spectrum core centered on the 
optical nucleus has been known about since the first VLA surveys 
of Seyferts
(e.g. Ulvestad \& Wilson 1984).
However, more recent, deeper maps at 20 and 6~cm have also revealed 
a series of ``blobs'' forming an elongated, curved structure extending 
out to $\sim$20~arcsec ($\sim 3.5 h_{75}^{-1}$~kpc) to the NE
(Baum et al. 1993; 
Miyaji, Wilson \& P\'{e}rez-Fournon 1992).
A much weaker, radio counter-structure to the SW of the nucleus has 
also been suggested (Wrobel \& Heeschen 1988) but, 
as yet, has not been confirmed.

In the optical and UV, regions of line emission have been detected
out to $\sim30$~arcsec ($\sim 5h_{75}^{-1}$~kpc) from the nucleus, 
with an asymmetric (often called Z-shaped) morphology
within the central 10~arcsec ($\sim 1.7h_{75}^{-1}$~kpc) -- see 
Ferruit, Wilson, Mulchaey (1998), and references therein.
The kinematic studies of this 
extended narrow emission line region (ENLR)
show peculiar motions
(e.g. Ulrich, P\'{e}quignot 1980)
while the stellar 
velocity field appears normal
(Arribas et al. 1997).
It has been 
suggested that the morphology and kinematics of the ionized gas
are the result of a bent, bipolar outflow from the nucleus
or due to entrainment within a precessing radio jet
(e.g. Veilleux, Tully, Bland-Hawthorn 1993; Ferruit et al. 1998).

X-ray emission from NGC~3516 was first detected in 1979 by the
Imaging Proportional Counter on board the  
{\it Einstein Observatory} (e.g. Maccacaro, Garilli \& Mereghetti 1987),
and studied by all the major X-ray observatories since.
The X-ray continuum exhibits spectral and temporal characteristics 
common to Seyfert 1 galaxies, including Fe K$\alpha$ emission,
absorption due to ionized gas (a ``warm absorber'') and an 
apparently variable power law continuum
(e.g. Guainazzi et al 2001, and references therein).
However, most relevant to the work discussed here, is the 
suggestion of extended emission in 0.1--2~keV band 
in data obtained using the 
High Resolution Imager (HRI) on board {\it ROSAT}.
Morse et al. (1995) found an elongation in X-ray emission 
on scales 10--30~arcsec (1.7--5$h_{75}^{-1}$~kpc)
approximately aligned with ENLR described above. 
However, as noted by Morse et al., this elongation could be
the result of residual errors in the spacecraft aspect solution.
The morphology of any extended X-ray emission within 
10~arcsec could not be studied in these HRI data
since any such emission was swamped by the 
instrumentally-scattered emission from the nucleus.

Here we present new X-ray data from NGC~3516 obtained 
with high spatial-resolution using the 
{\it Chandra X-ray Observatory} ({\it CXO}).
We restrict our discussion to the non-nuclear emission.
The results from the nuclear emission is presented in 
Netzer et al. (2002) and Turner et al. (in preparation).
In \S\ref{Sec:Obs} we describe the 
observations, and 
in \S\ref{Sec:spatial} the basic results from our spatial analysis.
The characteristics of an
off-nuclear source are described in 
\S\ref{Sec:serendip}, 
and those of the extended X-ray emission in 
\S\ref{Sec:xtended}.
We discuss our findings and present our conclusions in 
\S\ref{Sec:disc}.
Throughout we assume a Galactic column density of 
$N_{H I}^{gal} = 2.9\times10^{20}\ {\rm cm^{-2}}$ 
appropriate for this line of sight (Murphy et al 1996).


\section{THE OBSERVATIONS \& DATA REDUCTION}
\label{Sec:Obs}

\noindent
The data from NGC~3516 reported here were obtained from three 
observations using the {\it CXO}: a 47~ks observation 
in 2000 September, and two observations (of approximately 
40~ks and 75~ks respectively) in 2001 April 
(see Table~1).
All observations were performed with the ``S-array'' of the 
Advanced CCD Imaging Spectrometer (ACIS: e.g. Nousek et al 1998)
in the focal plane (with a temperature of $-120^{\circ}$~C).
The Low Energy Transmission Grating Spectrometer 
(LETGS: e.g. Brinkman et al. 1997) was in the optical path during 
the earlier observation, while the 
High Energy Transmission Grating Spectrometer 
(HETGS: e.g. Markert et al. 1994) was in place during the 
latter observations. 
In all cases, the zeroth-order image of the nucleus of NGC~3516
was close to the optical axis of the telescope and
fell on the (back-illuminated) ACIS-S3 chip.
Here we restrict our analysis to the non-nuclear emission detected
using the zeroth-order from the respective gratings.
The results from the nucleus of NGC~3516 
are discussed in Netzer et al. (2002) and Turner et al. (in preparation).


For completeness, we review a number of general
characteristics associated with 
the configuration of the telescope and detector.
Firstly, the ACIS pixels are 0.492~arcsec on a side
($85 h_{75}^{-1}$~kpc at the assumed distance to NGC~3516).
The X-ray telescope 
has a FWHM of 0.84~arcsec at energies $\lesssim 6$~keV
(the presence of neither the LETGS nor the HETGS 
in the optical path is thought to degrade this significantly). Thus 
the zeroth-order image of bright sources will be 
``piled-up''\footnote{This is the result of more than one incident 
	photon depositing its energy in a given pixel 
        within a single CCD frame. 
        The charge clouds from the individual photons 
        will be combined and hence interpreted by the on-board electronics 
	as due to a single  ``event'' by an incident photon of 
	higher energy.}
in their central pixel(s), and hence exhibit an artificial 
deficit of photons in the central pixel(s).
Secondly, the images of bright sources will also exhibit a 
read-out stripe\footnote{The read-out stripe is due to 
	events being detected by the CCDs while the image frame is being read 
	out. These ``out-of-time'' events are recorded at an incorrect row in 
	the CCD, and in the (RA,dec) coordinate system used here,
	will result in a stripe at an angle equal to the 
	roll angle of the spacecraft.}.
Thirdly, with the gratings in the optical path, the dispersed 
   ($1^{\rm st}$--, $2^{\rm nd}$--, $3^{\rm rd}$--order etc.)
   spectra will be projected onto the plane of the sky.
   For example, in the $1^{\rm st}$--order, the LETGS
   still has significant 
   effective area ($\sim 1\ {\rm cm^{-2}}$)
   at 9~keV. Such photons will 
   appear $\simeq$29~arcsec from the zeroth--order image.

In all subsequent analysis, the above effects have been taken 
into account as follows.
We ignore all events detected within 1~arcsec 
  (approximately 2~ACIS pixels) of the position expected (and detected) due 
  to the zeroth--order position of the target source 
  (i.e, the nucleus of NGC~3516). 
	``Pile-up'' is of negligible concern here 
	since we do not discuss the nuclear emission.
In addition, on an observation-by-observation basis, we give
 zero-weight to all events within $\pm 1$~pixel of 
  the read-out stripe.
Finally, outside a radius 25~arcsec we give
  zero-weight to all events 
within a polar angle $\pm 8$~degrees of the 
  location of the dispersed spectra.

The data analysis was performed using
the {\it Chandra} {\tt CALDB} (v2.7), and the 
{\tt CIAO} (v2.0.3) and {\tt HEAsoft} (v5.1) software packages.
Standard data reduction procedures were followed, 
including 
the removal of the ``streaks'' from the ACIS CCDs 
    (using {\tt destreak} v~1.3; Houck 2000), 
the selection of events with well-understood 
     ``grades''\footnote{Each photon detected 
	is assigned a ``grade'' based on the distribution of the charge 
	deposited within a $3 \times 3$ array of CCD pixels.
        Only grades equivalent to
	``{\it ASCA} grades''  0, 2, 3, 4 and 6 are used here.},
and the correction of a known error in the aspect solution 
    for the first observation.

\section{SPATIAL ANALYSIS}
\label{Sec:spatial}

\begin{figure*}[t]
\begin{center}
\includegraphics[scale=1.0,angle=0]{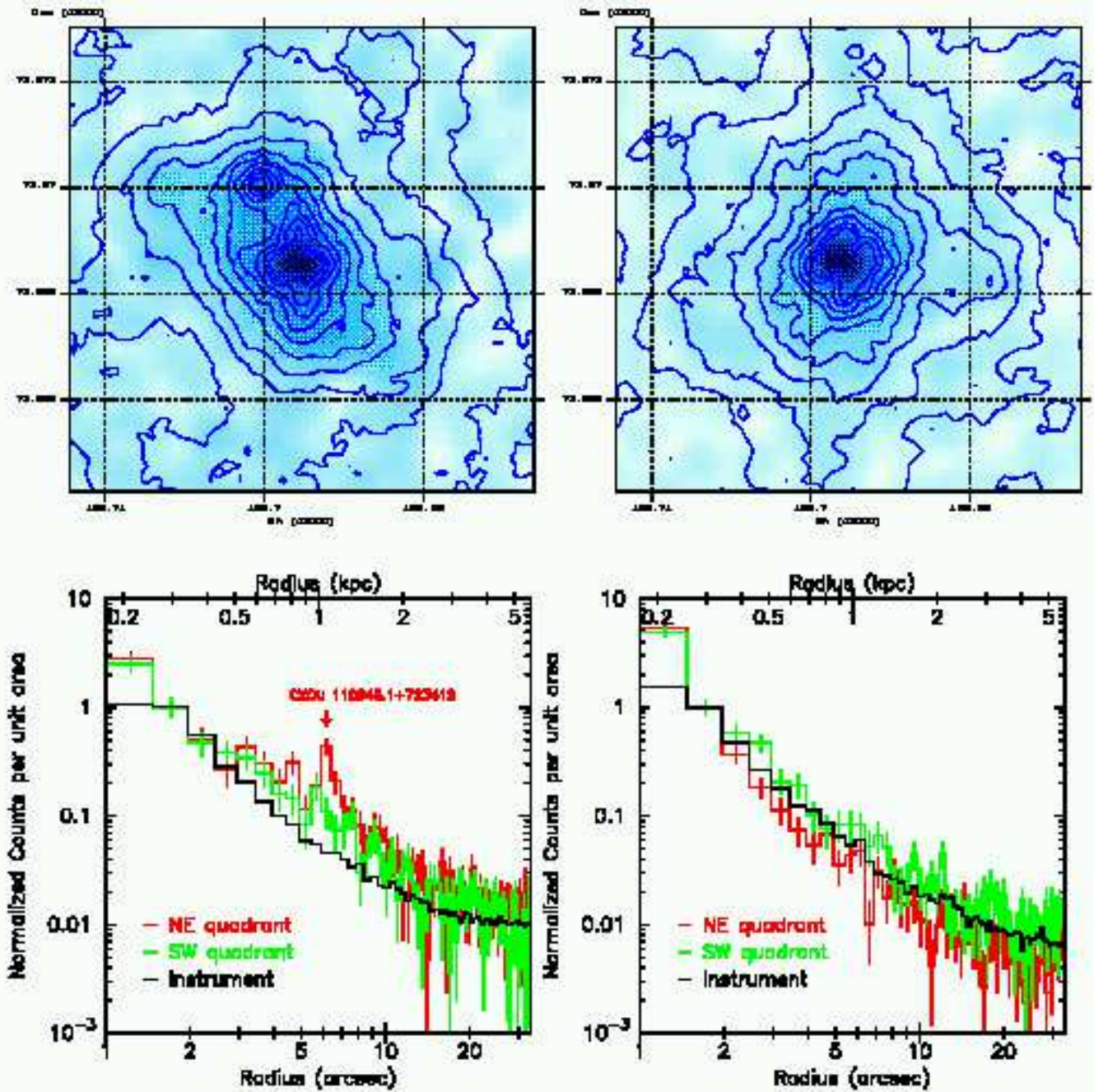}
\end{center}
\vspace{-1cm}
\caption{{\it Upper panels:} 
The central regions of NGC~3516 in the (left) 0.2--1~keV and 
(right) 4--8~keV bands. Each image is $64\times64$ pixels
centered on the optical nucleus, with pixels $0.492$~arcsec on a side
(corresponding to approximately $85 h_{75}^{-1}$~pc in NGC~3516). 
For clarity the image has been smoothed with a Gaussian with 
$\sigma =$3 pixels. The contours are derived from an adaptively smoothed 
version of the raw image.
{\it Lower panels:}
The corresponding radial point spread functions (rpsf)
from the NE (red) and SW (green) quadrants.
The black histograms show the rpsf derived from the observations of 
3C~273 (see text), and hence taken to represent that of the instrument.
\label{fig:spatial}}
\end{figure*}

\noindent
A bright point-like source is clearly evident in each observation
consistent (within the uncertainties of the {\it CXO} aspect reconstruction) 
with the position of the nucleus of NGC~3516
(RA=11h~06m~47.490s, dec=+72d~34m~06.88s, with a 95\% confidence 
radius of 0.25s; Clements 1981).
The residual errors in the aspect solutions (see Table~1)
were corrected manually, and the events then 
re-projected onto a common 
tangent-plane projection (centered on the optical nucleus).

In the upper panels of Fig.~\ref{fig:spatial} we show the images of the 
circumnuclear region in the 0.2--1~keV and 4--8~keV bands. 
The soft band image reveals an off-nuclear source 
$\sim$6~arcsec ($\sim 1.1 h_{75}^{-1}$~kpc)
to the NNE (position angle $\sim$29~degrees) 
of the nucleus of NGC~3516, and 
extended emission in the NE--SW direction stretching out to approximately
10~arcsec ($\sim 2 h_{75}^{-1}$~kpc) from the nucleus.
At high energies only the nucleus of NGC~3516 is 
detected, illustrating that neither the extended emission nor the 
off-nuclear source are obvious artifacts of the {\it CXO} 
aspect reconstruction.

In the lower panels of Fig.~\ref{fig:spatial} we show
the corresponding radial point spread functions (rpsfs) from the 
NE and SW quadrants.
For comparison also shown (the black histograms) are 
corresponding rpsfs from
the two archival HETGS observations of the bright quasar 3C~273 
currently available to us, superimposed on the the ambient background.
(The known X-ray jet in 3C~273 was excluded from the calculation.)
In the 0.2--1~keV band, excess emission is clearly seen in NGC~3516
in both the NE and SW quadrants at radii $\sim$3--10~arcsec
($\sim$0.5--1.7 $h_{75}^{-1}$~kpc). The spike in the rpsf of the 
NE quadrant at $\sim$6~arcsec is due to the off-nuclear source.
In contrast, the rpsf in the 4--8~keV band are consistent with the 
instrumental profile
(right panel of  Fig.~\ref{fig:spatial}).

\subsection{Extraction Regions}
\label{Sec:cells}

\begin{figure*}[t]
\vspace{-5cm}
\includegraphics[scale=1.0,trim=72 144 0 144, clip,angle=0]{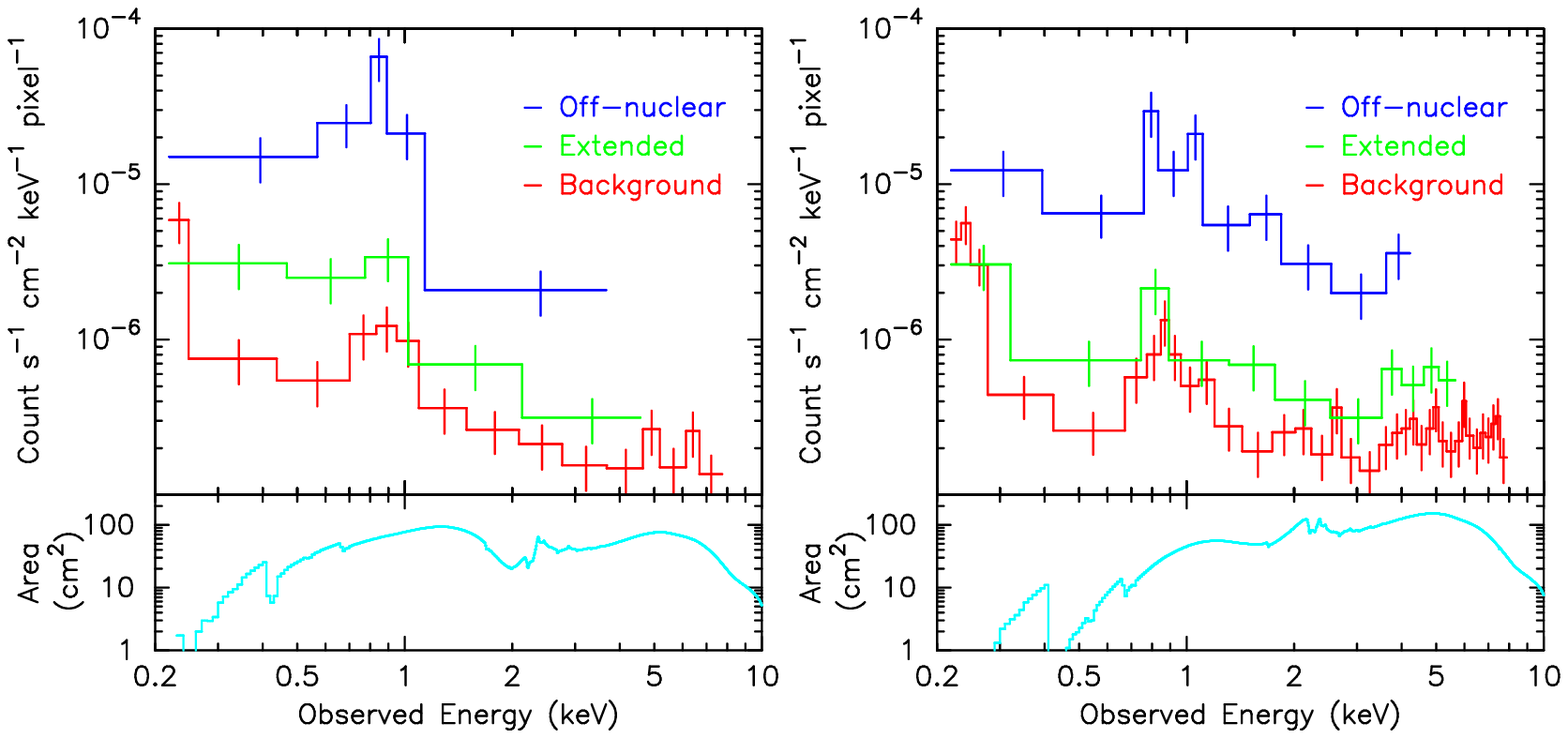}
\vspace{-6cm}
\caption{The spectra from three regions described in \S\ref{Sec:cells} 
derived from the LETGS (left) and HETGS (right). 
That from the off-nuclear source cell is shown in blue, the extended 
cell in green, and the background cell in red.
In each case the spectra are shown {\it per pixel}, and have been 
rebinned such that each bin contains at least 10~counts.
The lower panels show the effective area of each mirror/grating/detector
combination. 
\label{fig:spec_perpix}}
\end{figure*}

\noindent
For the subsequent analysis we make use of data extracted using the 
following regions.
The first region is a circle of radius 1.8~arcsec centered on the 
off-nuclear source (ie. equivalent to 42 ACIS pixels).
Hereafter we refer to this 
region as the ``off-nuclear source cell''.
The second region encompasses the extended emission, namely 
the NE and SW quadrant of an annulus stretching from 
3--10~arcsec from the nucleus of NGC~3516 (excluding the 
region around the off-nuclear source). 
This region encompasses 290 ACIS pixels, and hereafter we refer to 
it as the ``extended cell''.
The third region is used as a background region and consists of  
the NE and SW quadrant of an annulus stretching from 
10-20~arcsec from the nucleus.
This region encompasses 1545 ACIS pixels, and 
hereafter referred to  
as the ``background cell''.

In Fig.\ref{fig:spec_perpix} we show the counts spectra 
{\it per pixel} from each of these regions.
It should be noted that due to the different transmission
characteristics of the LETGS and HETGS, the effective area is different 
for each grating/detector combination. These are shown in the 
lower panels of Fig.\ref{fig:spec_perpix}, and necessitate 
separate spectral analysis of the two datasets.

\section{THE OFF-NUCLEAR SOURCE CXOU~110648.1+723412}
\label{Sec:serendip}

\noindent
Astrometry using the position of the optical nucleus of NGC~3516 gives a  
location of the off-nuclear source at (J2000)
RA=11h~06m~48.1s, dec=+72d~34m~12.5s, with an uncertainty of 
$\sim$1~arcsec. To the best of our knowledge, this is the first
detection of a distinct source at this location and we
designate it CXOU~110648.1+723412.

In the LETGS data set a total of 57 counts were detected in the 
0.2--8~keV band from the off-nuclear source cell. A total of 114 counts 
were detected in this band from this cell in the HETGS data set.
The ambient backgrounds in this band, derived using the 
extended cell (renormalising by the ratios of the areas of the cells)
are 8 counts in the LETGS and 20 counts in the HETGS data sets.

As illustrated in  Fig.\ref{fig:spec_perpix},
CXOU~110648.1+723412 is significantly detected in the $\sim$0.3--4~keV band
in both the LETGS and HETGS data sets.
However, with so few net counts, clearly 
only limited temporal and spectral 
information is available. 
We have grouped the spectra such that each new bin 
contained at least 20 counts, hence allowing an analysis 
using $\chi^2$-statistics. This resulted in only 2 bins for the 
LETGS data set, and 5 bins for the HETGS data set.
We find the spectra at all epochs to be consistent with 
a simple power-law continuum, a thermal bremsstrahlung continuum,
or emission from a collisionally ionized plasma.
For the power-law continuum, for which we obtain a best-fit with
a reduced $\chi^2$-statistic, $\chi^2_{\nu}$ = 1.90 
for 5 degrees of freedom ($dof$), the 90\% confidence ranges 
for the  photon index $(\Gamma$),
flux and luminosity in the 0.4--2~keV band (the latter 
after correcting for $N_{H I}^{gal}$, assuming isotropic emission 
and a location within NGC~3516) 
are $1.8 \lesssim \Gamma \lesssim 2.6$, 
$f$(0.4-2~keV) 
 $\simeq 2.1$--$3.4 \times 10^{-14}\ {\rm erg\ cm^{-2}\ s^{-1}}$,
and 
 $L$(0.4-2~keV)
 $\simeq 3.7$--$5.9 \times 10^{39}h_{75}^{-2}\ {\rm erg\ s^{-1}}$.
For the thermal bremsstrahlung continuum
($\chi^2_{\nu}/dof$ = 1.99/5), we find the temperature 
in the range $0.9 \lesssim kT$(keV)$ \lesssim 3.3$, 
$f$(0.4-2~keV) 
 $\simeq 1.8$--$4.3 \times 10^{-14}\ {\rm erg\ cm^{-2}\ s^{-1}}$,
and 
 $L$(0.4-2~keV)
 $\simeq 3.2$--$7.5 \times 10^{39}h_{75}^{-2}\ {\rm erg\ s^{-1}}$.
For a collisionally ionized plasma 
($\chi^2_{\nu}/dof$ = 2.24/4), we require the abundance of all 
elements to be $\lesssim 63$\% of their cosmic values, 
the temperature 
in the range $0.7 \lesssim kT$(keV)$ \lesssim 3.1$, 
$f$(0.4-2~keV) 
 $\simeq 1.4$--$4.8 \times 10^{-14}\ {\rm erg\ cm^{-2}\ s^{-1}}$,
and 
 $L$(0.4-2~keV)
 $\simeq 2.4$--$8.3 \times 10^{39}h_{75}^{-2}\ {\rm erg\ s^{-1}}$.
For these models, we find no requirement for absorption in excess of 
$N_{H I}^{gal}$, although the constraints that can be 
placed on any such component are poor
(typically $\lesssim$few$\times10^{21}\ {\rm cm^{-2}}$).
For future reference, at 90\% confidence we find an upper limit 
of the flux in the 4--8~keV band of 
$4\times10^{-14}\ {\rm erg\ cm^{-2}\ s^{-1}}$.

It should be noted that the above results 
do not change significantly if the 
background from the background cell is used instead
of that from the extended cell.
Such an approach would be more appropriate if the 
emission from within the extended cell was the result of 
a relatively few faint, discrete sources rather than 
truly-uniform, extended emission.

It is possible that CXOU~110648.1+723412 is extended.
Indeed our spatial analysis reveals evidence for 
some extension
on a scale $\sim$0.5~arcsec ($\sim0.2 h_{75}^{-1}$~kpc).
However given the co-addition of several observations, and the
likely presence of (probably patchy) 
underlying extended emission, 
further observations are required to investigate 
this possibility.

\section{THE EXTENDED EMISSION}
\label{Sec:xtended}

\noindent
The region used to extract the extended emission 
consists of 290 ACIS pixels ($70.2\ {\rm arcsec^{2}}$)
in the NE and SW quadrant of an annulus stretching from 
3--10~arcsec from the nucleus of NGC~3516
(excluding the region around CXOU~110648.1+723412).
Unfortunately there are insufficient counts per pixel 
to determine whether the X-ray emission from this region is 
smooth, patchy, and/or due to a relatively small
number of discrete sources. 
Thus below we discuss the ``mean'' X-ray characteristics of this region.

Totals of 55 and 141 counts were detected in the 
0.2--8~keV band from this region in the LETGS and HETGS
data sets (respectively). 
Renormalising for the areas, from the background cell 
we estimate ambient backgrounds of 
29 and 76 counts in the extended cell (respectively).
As illustrated in  Fig.\ref{fig:spec_perpix},
extended emission is significantly detected in the $\sim$0.3--4~keV band
in the LETGS data set, 
and at most energies in the $\sim$0.3--6~keV band in the HETGS data set.
However, as for CXOU~110648.1+723412, there are again
clearly too few net counts for detailed temporal or  spectral 
analysis. 
Grouping the spectra such that each new bin 
contained at least 20 counts results in  2 bins for the 
LETGS data set, and 7 bins for the HETGS data set.
We find the spectra to be formally consistent with a power law 
continuum ($\chi^2_{\nu}/dof$ = 0.83/7), 
with 90\% confidence ranges of 
    $0.6 \lesssim \Gamma \lesssim 2.8$, 
$f$(0.4-2~keV) 
 $\simeq 0.4$--$1.8 \times 10^{-14}\ {\rm erg\ cm^{-2}\ s^{-1}}$,
and 
 $L$(0.4-2~keV)
 $\simeq 0.8$--$3.1 \times 10^{39}h_{75}^{-2}\ {\rm erg\ s^{-1}}$.
A thermal bremsstrahlung continuum is also acceptable
($\chi^2_{\nu}/dof$ = 0.97/7), with
$kT$(keV)$ \gtrsim 1.3$, 
$f$(0.4-2~keV) 
 $\simeq 0.5$--$2.0 \times 10^{-14}\ {\rm erg\ cm^{-2}\ s^{-1}}$,
($0.7$--$2.8 \times 10^{-16}\ {\rm erg\ cm^{-2}\ s^{-1}\ arcsec^{-2}}$),
and 
 $L$(0.4-2~keV)
 $\simeq 0.8$--$3.3 \times 10^{39}h_{75}^{-2}\ {\rm erg\ s^{-1}}$
($1.1$--$4.8 \times 10^{37}h_{75}^{-2}\ {\rm erg\ s^{-1}\ arcsec^{-2}}$).
Clearly a number of more complex spectral forms are 
also consistent with the data.
We find no requirement for absorption in excess of 
$N_{H I}^{gal}$, although again the constraints that are placed on any such 
component are poor with the current data
($\lesssim$few$\times10^{21}\ {\rm cm^{-2}}$).
The current data have too low a 
signal-to-noise ratio to determine 
whether the spectra from the NE and SW quadrants differ
(e.g. whether they suffer different amounts of intrinsic absorption
as might be expected if they lie on different sides of the galactic
plane).
As for CXOU~110648.1+723412, at 90\% confidence the upper limit 
of the flux in the 4--8~keV band is
$4\times10^{-14}\ {\rm erg\ cm^{-2}\ s^{-1}}$.

\section{DISCUSSION \& CONCLUSIONS}
\label{Sec:disc}

\begin{figure*}[t]
\includegraphics[scale=1.0,angle=0]{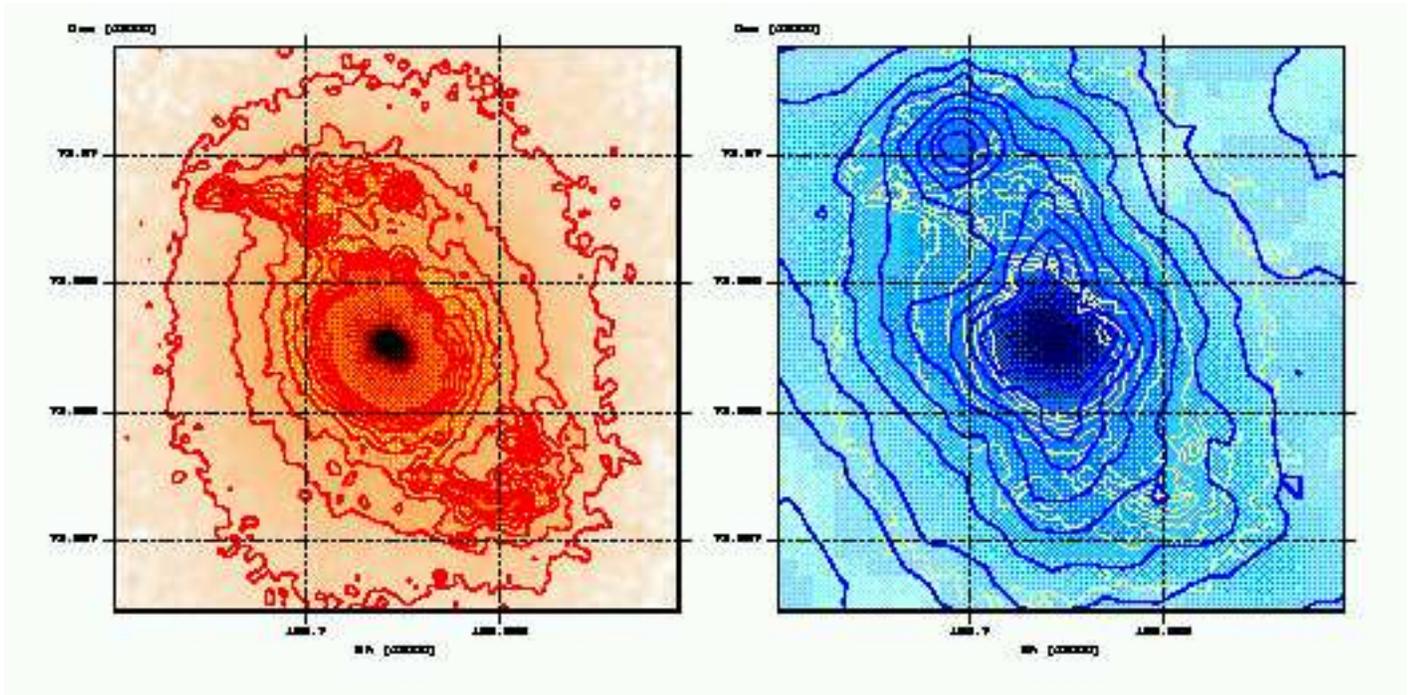}
\vspace{-1cm}
\caption{{\it Left:} {\it HST}/WFPC2 image of the circumnuclear 
regions of NGC~3516 in [O{\sc iii}] ($5007$\AA).
The image is 15.7~arcsec on a side,  
corresponding to $2.73 h_{75}^{-1}$~kpc at the distance of NGC~3516.
Linearly-spaced contours are overlaid to illustrate 
the structures to the NE and SW of the nucleus.
{\it Right:} The inner regions of the 0.2--1~keV image from Fig.~1
with the morphology of the  
[O{\sc iii}] emission shown by the 
white contours (only every other contour is plotted for clarity).
\label{fig:uv}}
\end{figure*}

\noindent
The spatial resolution of {\it CXO} finally allows the innermost regions 
($\lesssim$1~kpc) of the nearest galaxies to be probed in X-rays.
Already observations have shown that besides point sources 
(many of which appear to be emitting at super-Eddington luminosities),
a large fraction of galaxies have extended
X-ray emission on scales $\sim$0.5--1.5~kpc 
(see Weaver 2001 for a recent review).
Furthermore, this extended X-ray emission often has an anisotropic 
morphology, and is aligned with other anisotropic structures 
such as the radio emission and/or ENLR.

The data presented here show that NGC~3516 fits into this picture.
We have found a distinct off-nuclear source (CXOU~110648.1+723412) 
which is embedded within the extended X-ray emission
$\sim 6$~arcsec ($1.1 h_{75}^{-1}$~kpc) NNE of the nucleus.
More interestingly, 
we have found extended X-ray emission in the $\sim$0.3--4~keV band, 
detected out to a radius $\sim 10$~arcsec ($\sim 2 h_{75}^{-1}$~kpc)
from the nucleus. 
The extended emission 
appears to be anisotropically distributed in the NE--SW direction, 
although instrumental artifacts (most notably the read-out streak
associated with the relatively-bright nuclear emission) prevent us
from being certain.
However we do note that an elongation in this direction 
was suggested by Morse et al. (1995) based on {\it ROSAT} HRI data.
In the left panel of 
Fig.~\ref{fig:uv} we show the circumnuclear regions of NGC~3516
in the optical band, obtained from an archival {\it HST}/WFPC2 
observation using the linear ramp filter (FR533N)
centered on the red-shifted [O{\sc iii}] ($5007$\AA) emission line.
As shown previously by Ferruit et al. (1998), the morphology of 
the [O{\sc iii}] emission is complex, but shows a distinct 
anisotropy in the  NE--SW direction. 
The right panel of Fig.~\ref{fig:uv} compares the 
[O{\sc iii}] and soft X-ray morphologies. 
Clearly there appears to be good agreement between the alignments.

\subsection{The Off-nuclear source}
\label{Sec:disc-serendip}

\noindent
The source CXOU~110648.1+723412 is highly likely to be 
located within the galaxy NGC~3516. 
The density of background active galactic nuclei (AGN) 
at this flux level in the 0.5--2~keV band 
is $\sim20$ per square degree (e.g. Hasinger et al. 1998), thus 
the probability of a background source within 6~arcsec of the 
nucleus of NGC~3516 is $\sim2\times10^{-4}$.
Similarly the probability that the emission is due to a foreground 
(Galactic) source is at least a factor 2 lower 
(e.g. Kuntz \& Snowden 2001).

As stated in \S\ref{Sec:serendip}, based on the data currently 
available we cannot determine unambiguously
whether CXOU~110648.1+723412 is a true point-source, or a relatively
small, but diffuse, enhancement in the extended emission.
If it is a point-source, indeed located in NGC~3516 and 
emitting isotropically, then its X-ray luminosity 
($L(0.4-2$~keV$) \sim$2--8$\times10^{39}h_{75}^{-2}\ {\rm erg\ s^{-1}}$)
is above the Eddington luminosity for a $1.4 M_{\odot}$ 
neutron star. Such ``super-luminous'' off-nuclear sources are being 
regularly found by {\it CXO}
(e.g. Griffiths et al. 2000; Fabbiano, Zezas, Murray 2001; 
George et al 2001; Pence et al 2001)
with similar spectral characteristics as CXOU~110648.1+723412.
Alternatively, CXOU~110648.1+723412 could be the result of 
the superposition of several sub-Eddington sources, or simply 
a local enhancement within the diffuse emission.
          Unfortunately, given the FWHM of the {\it ROSAT} HRI 
          ($\sim$ 4 arcsec), plus the additional blurring due to errors 
          in the aspect solution, the X-ray emission detected by the HRI 
          at the location of CXOU~110648.1+723412 is dominated by that from 
          the nucleus itself (e.g. see Morse et al. Fig 3).
          Thus, no useful constraints 
          can be placed on the long-term variability characteristics 
          CXOU~110648.1+723412 at this time.
Higher-quality X-ray spectra and temporal studies are required to 
determine the origin of this emission.

Interestingly, as can be seen from the 
right panel of Fig.~\ref{fig:uv}, CXOU~110648.1+723412
is located just N of the [O{\sc iii}] E--W linear structure 
to the NE of the nucleus.
The lack of  [O{\sc iii}] emission in the immediate vicinity 
of the X-ray source may be the result of its radiation field 
photo-ionizing oxygen to O{\sc iv} and higher.

\subsection{The Extended X-ray Emission}
\label{Sec:disc-xtended}

\noindent
There are several possible origins for the extended 
X-ray emission reported here. Clearly it might 
not be truly extended emission at all, instead being 
the result (at least in part) of a relatively-small number of 
faint, discrete sources. Certainly we see 
one clear discrete source, CXOU~110648.1+723412, 
and there are some indications of patchiness within the 
rest of the extended emission. The current {\it CXO} data simply 
do not have sufficient signal-to-noise to 
usefully constrain the possibility.
Nevertheless, if the observed X-ray emission is (primarily)  
extended, then it may be the result of 
emission from  hot, collisionally-ionized gas, 
perhaps related to the presence of active 
star-forming regions.
Alternatively, the extended X-ray emission may be produced by gas 
photo-ionized by the intense nuclear radiation field, and include the 
electron-scattering of the nuclear radiation.
Such gas may be related to the bulk outflow known to 
exist in this source (e.g. Hutchings et al. 2001, and references therein), 
perhaps as a result of a ``wind'' from the inner regions
(e.g. Pietrini, Torricelli-Ciamponi 2000; 
Krolik, Kriss 2001, and references therein).
Of course at least some of the gas could be under the influence of 
both collisional- and photo-ionization processes.
In principle the dominant ionization mechanism 
can be determined from a detailed study of the X-ray spectrum. 
Unfortunately the current {\it CXO} data simply 
do not have sufficient signal-to-noise.

Diffuse emission in the 0.4--2~keV band, with a luminosity
$\lesssim$few$\times10^{39}\ {\rm erg\ s^{-1}}$, has been detected 
from the inner few kpc of a number of galaxies.
These include ``classic'' AGN
  (e.g.  the Seyfert 1.5 galaxy NGC~4151; Ogle et al. 2000, and 
         the Seyfert 2 galaxies Circinus and NGC~1068;
             Sambruna et al. 2001; Young, Wilson, Shopbell 2001),
ultraluminous infrared galaxies 
	(e.g. Mrk~273; Xia et al 2001),
and galaxies with low-luminosity AGN 
	(e.g. M51; Terashima \& Wilson 2001).
Interestingly, however, extended X-ray emitting gas with similar 
characteristics has also been observed in a number of 
``normal'' galaxies 
       (e.g. the starburst galaxy M82; Grifffiths et al. 2000, 
             the nearby spiral M81; Tennant et al. 2001, 
             and the Sa galaxy NGC~1291; Irwin, Sarazin \& Bregman 2001).
Thus it is not clear how (if at all) the diffuse X-ray emission
in NGC~3516 might be related to the presence of an active nucleus.
Nevertheless it is of interest to explore the case 
where the extended emission arises solely as a result of
electron-scattering  of the nuclear radiation 
field.
If we assume that the electrons are distributed in clouds, 
which have a volume filling factor, $f_V$, then 
for a radius $r$, the luminosity of the scattered radiation 
is given by $L_{scat} \simeq L_{pri} \sigma_{\rm T} f_V n_e r$,
where 
$L_{pri}$ be the mean (time-averaged) luminosity of the nuclear source, 
and $\sigma_T$ the Thomson cross-section. 
The luminosity of the nucleus is known to be variable. 
Here we simply assume the approximate average luminosity of 
the observations made to date, namely
$L_{pri}$(0.4--2~keV) 
	$\simeq 4\times10^{42} h_{75}^{-2}\ {\rm erg\ s^{-1}}$. 
From \S\ref{Sec:xtended}, 
$L_{scat}$(0.4--2~keV) 
	$\sim 2\times10^{39} h_{75}^{-2}\ {\rm erg\ s^{-1}}$, 
thus for $r=1$~kpc, 
$f_V n_e \sim 0.2\ {\rm cm^{-3}}$. 
The X-ray ionization parameter
(defined over the 0.1--10~keV band) is 
$U_X \sim 0.05 f_V$
and the column density (out to 2~kpc) is 
$N_{\rm H} \sim 7\times10^{20}\ {\rm cm^{-2}}$.
Unless $f_V \lesssim 10^{-2}$,
this combination of $U_{X}$ and
$N_{\rm H}$ imply the gas is indeed highly ionized
and hence our assumptions are valid.
The strongest spectral features emitted by such gas 
are likely to be
C{\sc vi}, O{\sc vii}, O{\sc viii}, and Ne{\sc ix} 
recombination continua and emission lines. We note that
similar features have been observed in X-ray spectra of Seyfert 2
galaxies (e.g., NGC~1068; Young, Wilson, \& Shopbell 2001). 
We also note that under the above assumptions, 
the extended gas has a significantly smaller column density 
than the ``warm-absorbers'' seen in Seyfert 1 galaxies
(typically 
$10^{21}$--$10^{23}\ {\rm cm^{2}}$, e.g. George et al. 1998).
Indeed, Netzer et al (2002) have shown that the warm-absorber
in NGC~3516 has a column density $\sim 10^{22}\ {\rm cm^{2}}$, 
a density $n_e \gtrsim$few$\times10^{6}\ {\rm cm^{-3}}$,
and is located at a radius $r \lesssim$0.5~pc from the nucleus.
Thus these data do not support the idea that the 
gas responsible for the warm absorbers seen in Seyfert 1 
galaxies is the same gas that is seen as extended 
X-ray emission 
in Seyfert 2 galaxies.

Finally, if the 
extended X-ray emitting gas is indeed photo-ionized, 
then for $f_V \simeq 1$
its temperature is $T \simeq 10^{5}\ {\rm K}$.
Thus $n_e T \sim$ few$\times10^{3}\ {\rm cm^{-3}\ K}$.
As implicit from \S\ref{sec:intro}, the extended X-ray emitting gas 
appears to be 
co-spatial with the ENLR in NGC~3516 (see also Fig.~\ref{fig:uv}).
Photo-ionization models of the optical line ratios observed from 
gas responsible for the ENLR give $n_e T \lesssim $few $\times10^{5}$--
few $\times10^{6}\ {\rm cm^{-3}\ K}$
(e.g. Ulrich \& P\'{e}quignot 1980; Aoki et al. 1994).
Hence the pressure ratio between the gas responsible for the 
extended X-ray emission and ENLR is $\lesssim 10^{-2}$. 
The two are not in pressure equilibrium so the  X-ray emitting gas
cannot be the confining medium for the ENLR clouds.
Such a situation has also been found in the case of NGC~4151
(Ogle et al. 2000).
There may well be other effects such as a dynamical interaction 
between these two gaseous components, but 
an exploration of this is beyond the scope of the current paper.


\acknowledgements

We thank Keith Arnaud for providing some of the software 
used for this project, and an anonymous referee for several useful
suggestions.
We acknowledge the financial support of the
Joint Center for Astrophysics (IMG, TJT), 
NASA (TJT, through grant number NAG5-7385),
and
the Universities Space Research Association (KN).
This research has made use of the Simbad database, operated by 
the Centre de Donn\'{e}es astronimiques de Strasbourg (CDS);
the {\tt VizieR} Service for Astronomical Catalogues,
developed by CDS and ESA/ESRIN;
the NASA/IPAC
Extragalactic Database (NED), operated by the Jet Propulsion Laboratory,
California Institute of Technology, under contract with 
NASA; 
and of data obtained through the HEASARC on-line service, 
provided by NASA/GSFC.

\typeout{REFERENCES}

\end{document}